\documentclass{elsart}
\usepackage{epsfig}
\begin{document}

\begin{frontmatter}

\title{Monte Carlo Study of Ordering and Domain Growth in a Class
of fcc--Alloy Models}

\author{M. Kessler, W. Dieterich}

\address{Fachbereich Physik, Universit\"at Konstanz\\
D-78457 Konstanz, Germany}

\author{A. Majhofer}

   \address{Institute of Exp. Physics,
Warsaw University \\ Hoza 69 PL-00681 Warszawa, Poland}

\begin{abstract}
Ordering processes in fcc--alloys with composition $A_3B$ (like
Cu$_3$Au, Cu$_3$Pd, CoPt$_3$ etc.) are investigated by Monte Carlo
simulation within a class of lattice models based on
nearest--neighbor (NN) and second-neighbor (NNN) interactions.
Using an atom--vacancy exchange algorithm, we study the growth of
ordered domains following a temperature quench below the ordering
spinodal. For zero NNN-interactions we observe an anomalously slow
growth of the domain size $L(t) \sim t^\alpha$, where $\alpha \sim
1/4$ within our accessible timescales. With increasing
NNN--interactions domain growth becomes faster and $\alpha$
gradually approaches the value 1/2 as predicted by the
conventional Lifshitz--Allen--Cahn theory.

\end{abstract}

\begin{keyword}
Ordering kinetics \sep domain coarsening \sep binary alloys \sep
Monte Carlo
\PACS 05.50.+q \sep 64.60.-i \sep 64.60.Cn
\end{keyword}

\end{frontmatter}

\section{Introduction}
Ordering transitions in real metallic alloys normally show kinetic
properties which are substantially more complex than the standard
Lifschitz--Allen--Cahn scenario in the case of a scalar,
non--conserved order--parameter \cite{Lifsh,AllenC}: i) The
symmetry of the ordered phase can imply a multicomponent
order--parameter and the appearance of several antiphase domains.
ii) There may be different types of domain boundaries providing
different forces for curvature--driven coarsening. iii) Coupling
of non--conserved order--parameter components to the (conserved)
alloy composition generally leads to compositional changes within
domain boundaries. Domain growth hence is accompanied and slowed
down by interdiffusion processes. iv) Depending on properties of
the chemical interactions, vacancies may enrich in the domain
boundaries and thus enhance the dynamics when atom--vacancy
exchange is the prevailing migration mechanism \cite{Porta,Fron}.
v) Quenching into a two--phase region in the
temperature--concentration plane releases competing processes of
ordering and spinodal decomposition, eventually accompanied by the
appearance of transient phases. vi) Finally we mention effects
caused by surfaces or imperfections, which break the lattice
translational symmetry.

Despite a large body of literature \cite{Ducast}, many questions
in this area have remained open. Our aim here is to study some new
qualitative aspects primarily in connection with the first two
points listed above. For that purpose we consider fcc $A_3B$--type
alloys (such as Cu$_3$Au, Cu$_3$Pd etc.) that display an ordered
$L1_2$--structure. A class of models is considered which includes
chemical interactions on the fcc--lattice up to second neighbors.
Our main result will be to demonstrate a gradual changeover from
an anomalously slow growth with an effective exponent $\alpha \sim
1/4$ to conventional growth, characterized by $\alpha = 1/2$, when
the second neighbor interactions are increased from zero. Results
will be interpreted in terms of properties of low--energy or
type--I domain walls, whose energy vanishes as the
second--neighbor interactions are switched off.

\begin{figure}[b]
\begin{center}
\epsfig{file=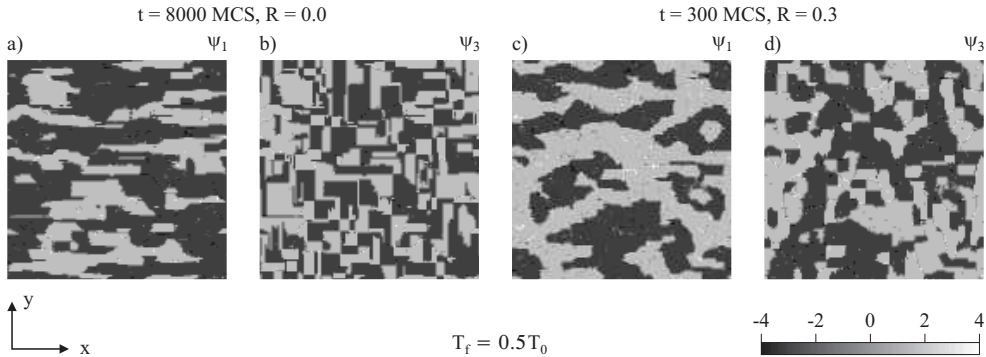, width=13cm} \caption{Comparison of domain
patterns reflected by $\psi_1$ at $T = 0.5\,T_0$ for a) $R = 0$;
$t = 8\cdot10^3$~MCS  and c) $R = 0.3$; $t = 3\cdot10^2 $~MCS,
while $\psi_3$--patterns are shown in b) and d) for the same
parameters. From the figures type--I and type--II walls can
clearly be distinguished.}
\end{center}
\end{figure}

\section{Fcc--alloy model with vacancy--driven kinetics}

Consider an fcc--lattice, where each site i can be occupied by an
$A$-atom, a $B$-atom or a vacancy ($V$). Occupation numbers thus
satisfy $c_i^A + c_i ^B + c_i^V = 1$. Their averages over all
sites are chosen in accord with stoichiometric $A_3B$ alloys, $c^A
= 3 c^B$. The average vacancy concentration $c^V \ll 1$ is taken
small enough so that static properties are unaffected by the
vacancies. For NN--sites $i$ and $j$ interactions are taken as in
our previous work \cite{paper2}, $V_{ij}^{BB} = V^{BB}
> 0$; $V_{ij}^{AA} = V_{ij}^{AB} = - V^{BB}$.
On the other hand, for NNN--sites $i$ and $j$ we assume
$V_{ij}^{BB} = V^{AA}_{ij} = R V^{BB}$; $V_{ij}^{AB} = 0$, with $0
\leq R \leq 0.5$.

The above model displays a 4-fold degenerate ground state
corresponding to the ideal $L1_2$--structure. This structure is
described by a 4--component order parameter \cite{Lai}, $\Psi =
(\psi_0, \psi_1, \psi_2, \psi_3)$, where $\psi_0$ refers to the
$A/B$--concentration, while $\psi_1$, $\psi_2$ and $\psi_3$
describe a succession of $B$-rich and $B$--depleted atomic layers
along the $x$--, $y$-- and $z$--direction, respectively. Ordered
domains are of the type $(\psi_1,\psi_2, \psi_3) \propto (-1,1,1);
(1,-1,1); (1,1,-1)$ and $(-1,-1-1)$. A sign--change of $\psi_1$
along the $x$--direction implies a high--energy type--II domain
wall. Along the $y$-- or $z$--direction a sign change of $\psi_1$
can be effected without braking nearest--neighbor bonds, and
creates a low--energy type--I wall. Its energy solely results from
NNN--interactions and therefore is zero if $R = 0$.

In our Monte Carlo simulations we employ the atom--vacancy
exchange algorithm. The ordering temperature $T_0$ and the
temperature for spinodal ordering $T_{sp}$ are deduced from
simulations as in \cite{paper1}. At $R= 0$ we have $k_B\,T_0 =
1.83\,V_{BB}/2$, while $T_{sp} \simeq 0.967\,T_0$. $T_0$ and
$T_{sp}$ increase nearly linearly with $R$, such that
$T_0(R)/T_0(0) \simeq 2.38$ and $T_{sp}(R)/T_{sp}(0) \simeq 2.43$
at $R=0.5$.

Typical domain patterns observed after a sudden quench from
infinite temperature to a final temperature $T < T_{sp}$ are shown
in Figs. 1a,b for $R = 0$ and in Figs. 1c,d for $R = 0.3$.
Evolution times after the quench were chosen such that the typical
domain sizes (see section 3) nearly agree in both cases. Apart
from the much faster evolution in the case $R = 0.3$ the most
important observation is that type--I walls in Figs. 1a,b are flat
and nearly perfect, whereas in Figs. 1c,d they show curvature and
larger fluctuations. A connection of these different behaviors to
kinetic properties will be discussed in section 3.

Using experimental tracer diffusion constants for calibration, our
algorithm allows us to relate the Monte Carlo time to the physical
timescale in kinetic processes. Details of this analysis, applied
to Cu$_3$Au, are described in \cite{paper2}. As an
order--of--magnitude approximation we find that $1$~MCS roughly
corresponds, for example, to $0.1s$ at $T=0.9\,T_0$ and $10^7s$ at
$T=0.5\,T_0$. We come back to this point in section 3 when
discussing different temporal regimes in domain coarsening
processes.

\begin{figure}[hb!]
\begin{center}
\epsfig{file=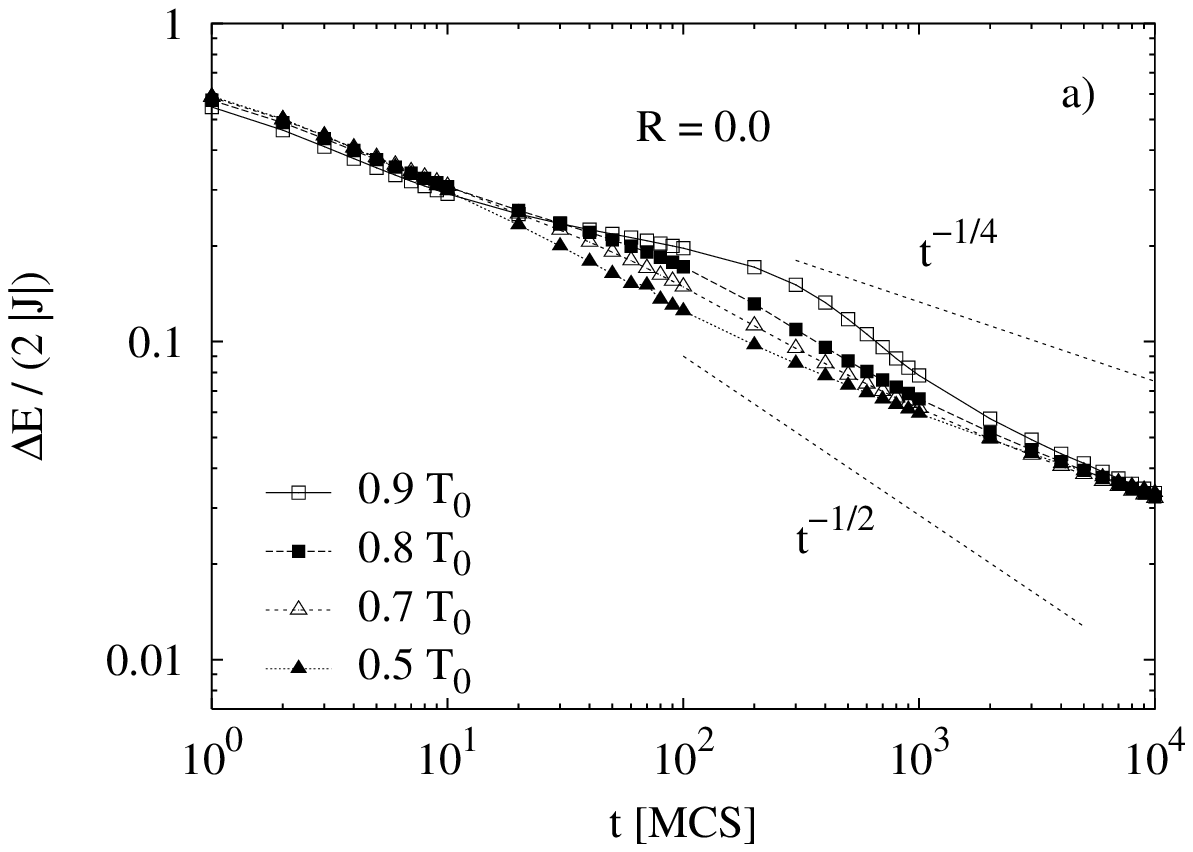, width=6.5cm} \epsfig{file=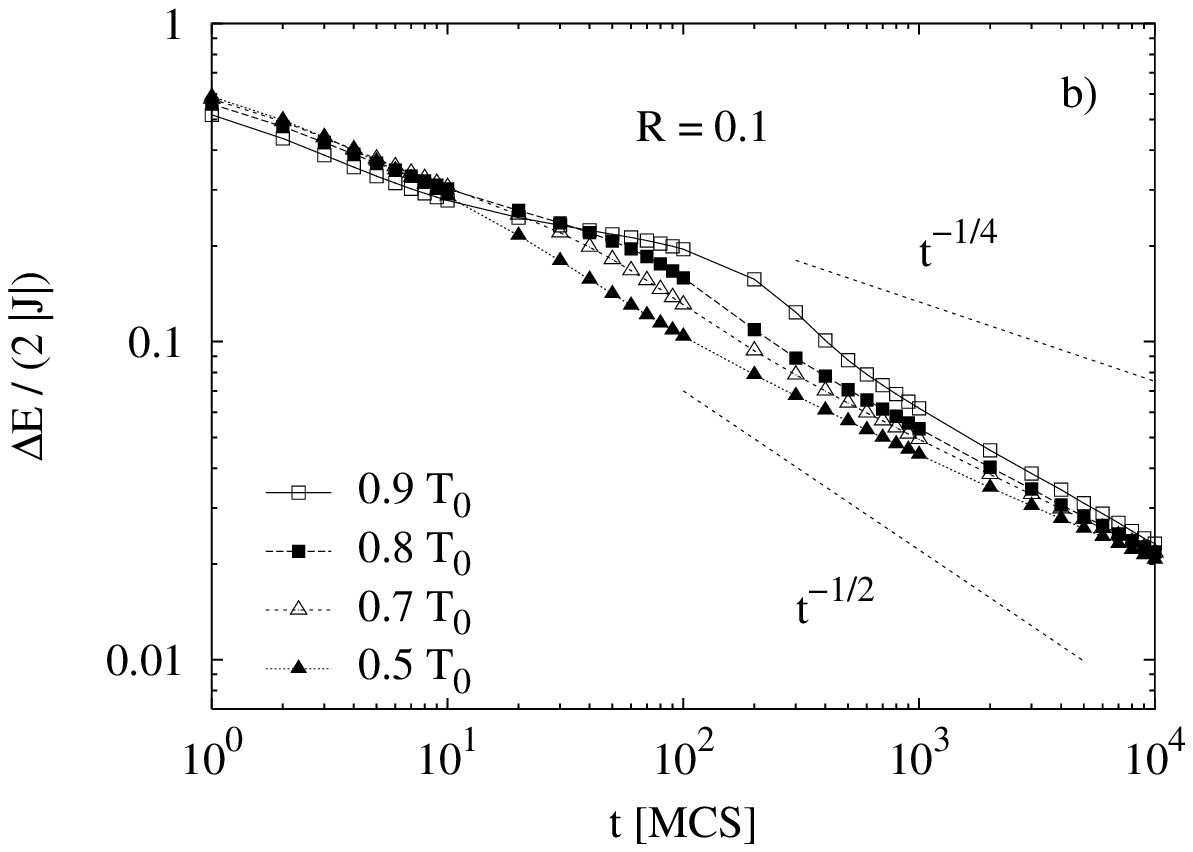,
width=6.5cm} \epsfig{file=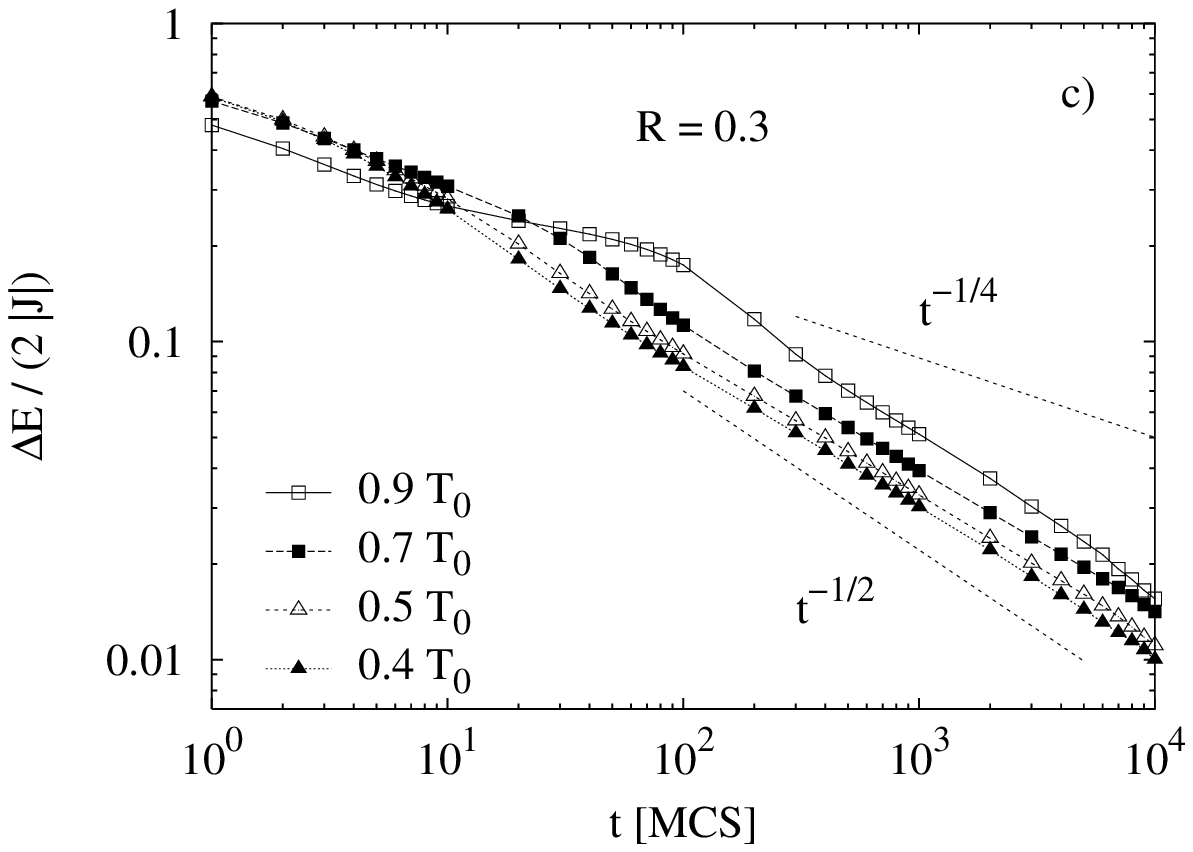, width=6.5cm}
\epsfig{file=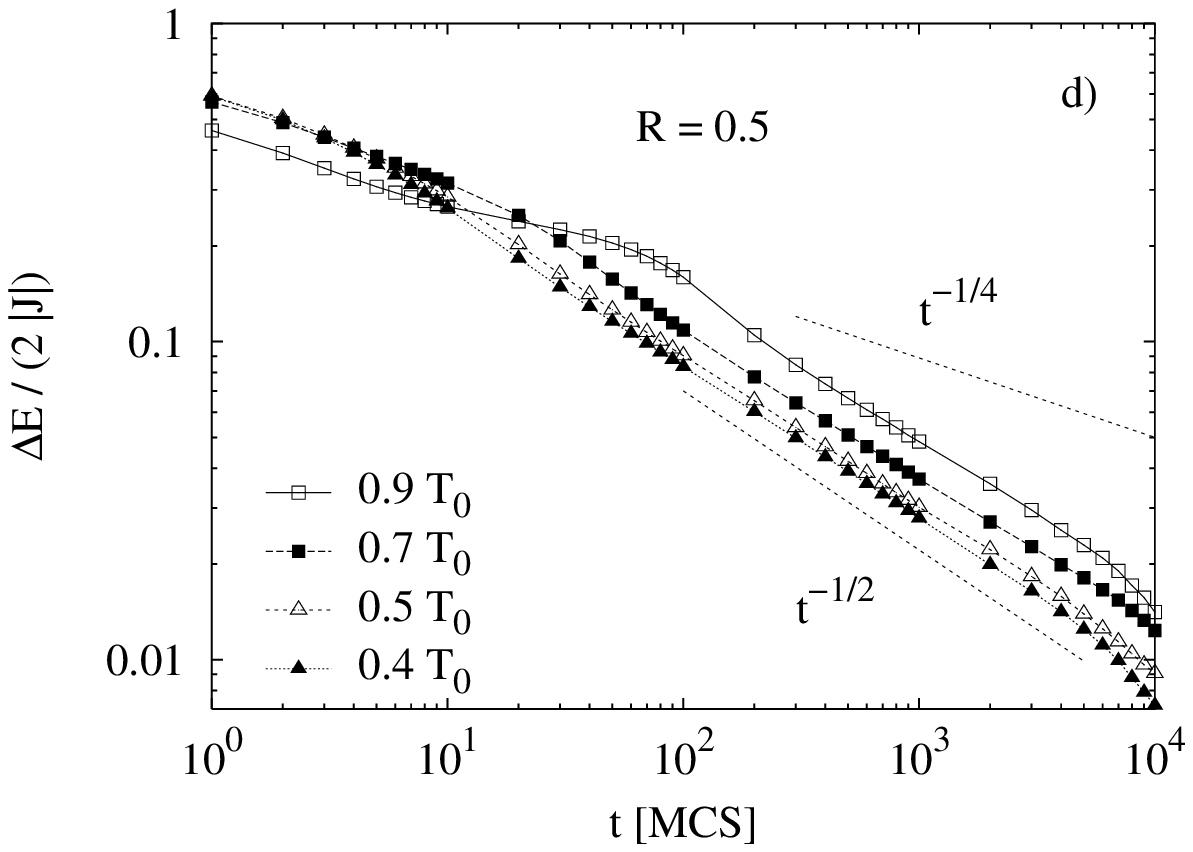, width=6.5cm} \caption{Evolution of the
excess energy $\Delta E(t)$ per lattice site for different
temperatures $T$ and different NNN--interactions: a) $R = 0$; b)
$R = 0.1$; c) $R = 0.3$ and d) $R = 0.5$. Dotted lines have slopes
corresponding to $\alpha = 1/4$ and $\alpha = 1/2$.}
\end{center}
\end{figure}

\section{Domain growth -- results and discussion}
For a quantitative analysis of the structure and growth of domains
we introduce equal--time structure factors $S_{\alpha}(\vec k, t)
= \langle |\Psi_\alpha (\vec k, t)|^2 \rangle$. In view of the
sign changes in the local order parameters  $\psi_\alpha(\vec r,
t)$ across a wall, see section 2, it is clear that the linear
combinations
\begin{equation}\label{5}
S_{||}(k,t) = \frac{1}{3} \left( S_1(k,0,0,t) + S_2(0,k,0,t) +
S_3(0,0,k,t)\right)
\end{equation}
\vspace{-0.7cm}and\vspace{-0.5cm}
\begin{equation}\label{6}
  S_\perp (k,t) = \frac{1}{N_k} \sum_{q_1^2+q_2^2=k^2}
  \left(S_1(0,q_1,q_2,t) +
  S_2(q_1,0,q_2,t)+S_3(q_1,q_2,0,t)\right)
\end{equation}
\vspace{-0.3cm} reflect the arrangement of type--II and type--I
walls, respectively. In Eq. (\ref{6}), $N_k$ denotes the number of
pairs $(q_1, q_2)$ that fulfill $q_1^2 + q_2^2 = k^2$. Typical
distances between walls of either type are determined by the first
moments $k_{||}(t)$ and $k_{\perp}(t)$ of the structure factors
(\ref{5}) and (\ref{6}). In addition we study the excess energy
stored in the domain walls, $\Delta E (t)  = E(t) - E(\infty)$,
where $E(\infty)$ is the energy after complete equilibration.

\begin{figure}[hb!]
\begin{center}
\epsfig{file=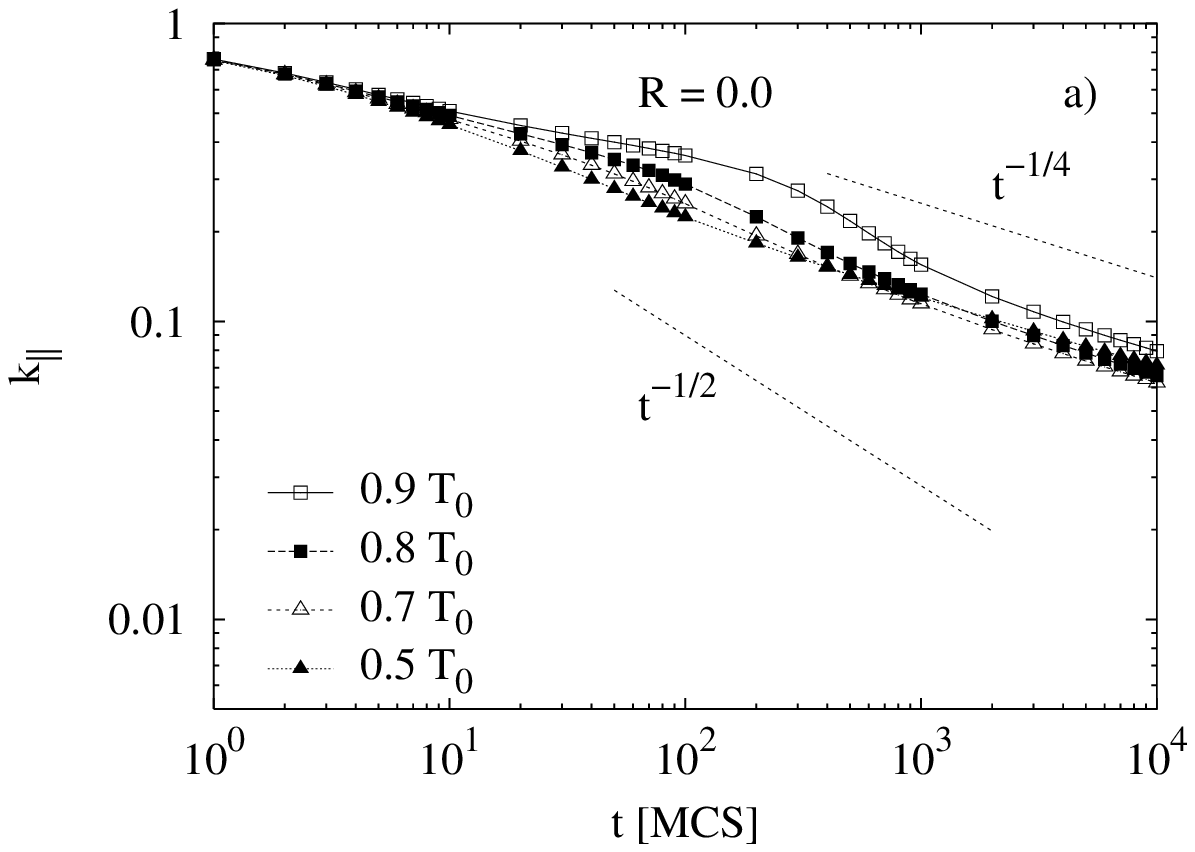, width=6.5cm} \epsfig{file=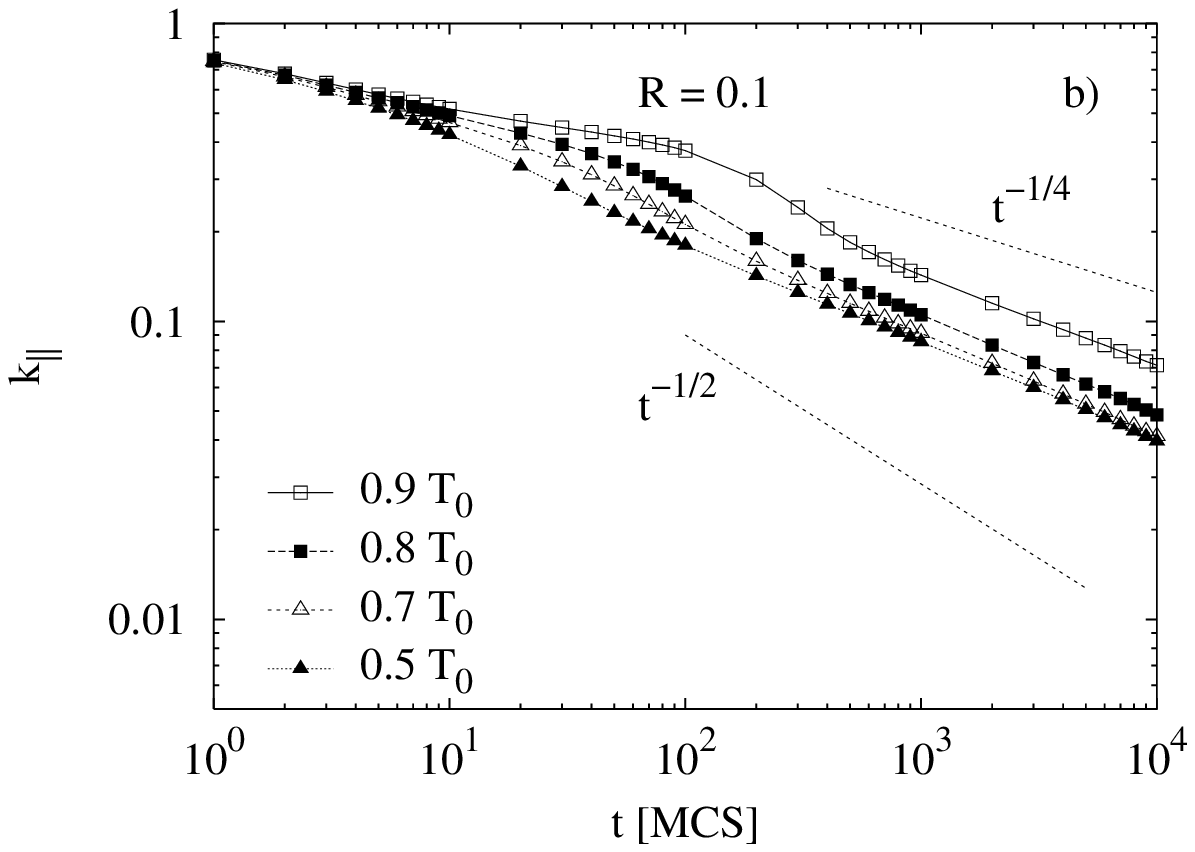,
width=6.5cm} \epsfig{file=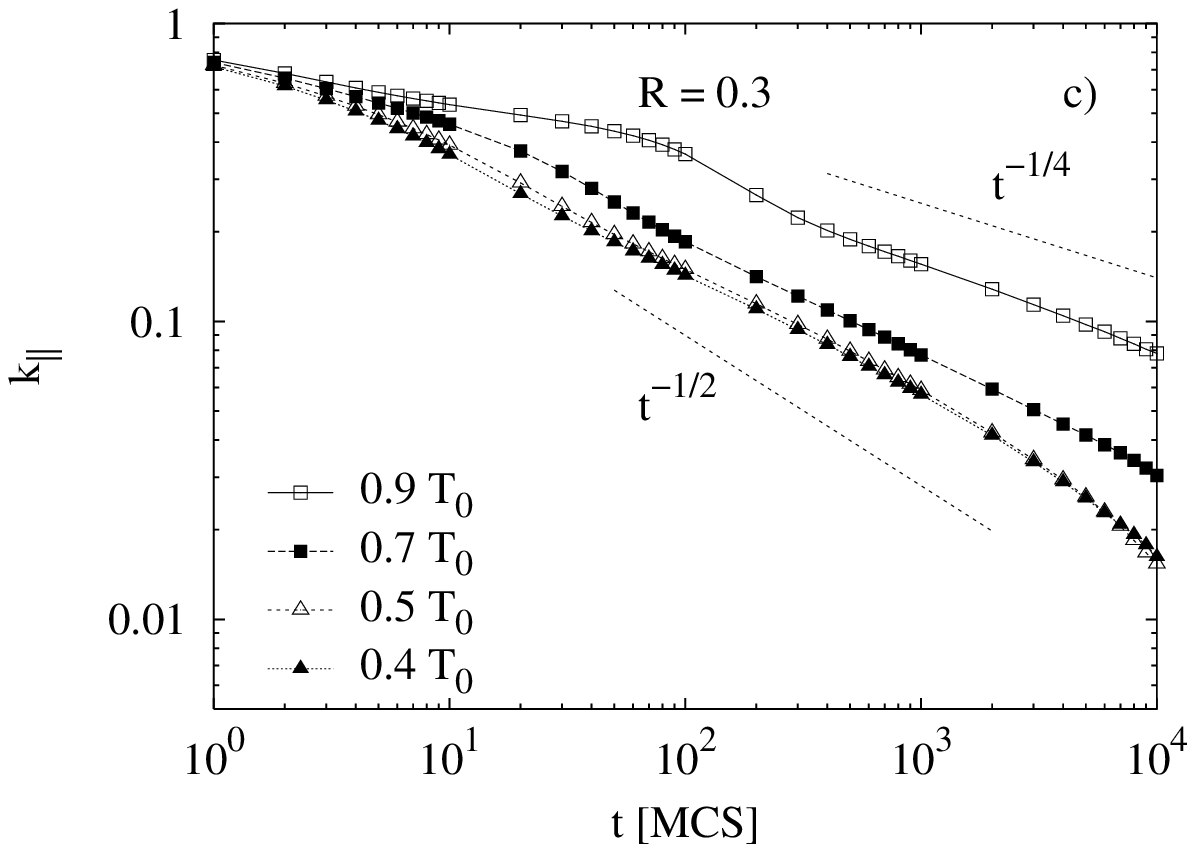, width=6.5cm}
\epsfig{file=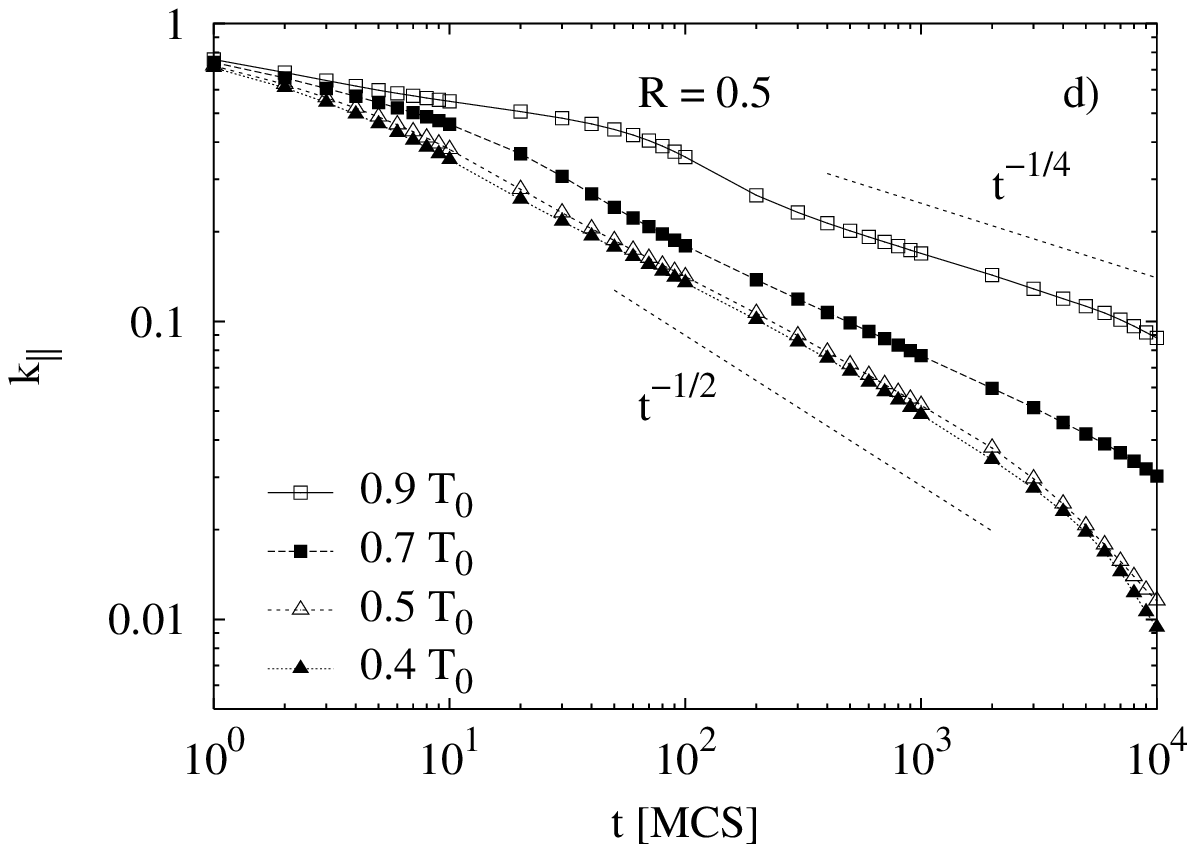, width=6.5cm} \caption{Evolution of the
first moment $k_{||} (t) $ of the structure factor $S_{||}(k,t)$
for different temperatures $T$ and different NNN--interactions: a)
$R = 0$; b) $R = 0.1$; c) $R = 0.3$ and d) $R = 0.5$.}
\end{center}
\end{figure}

In the following we focus on $T < T_{sp}$. In Fig. 2a-d $\Delta
E(t)$--data are presented for different $R$. At the highest
temperature $T = 0.9\, T_0$, which in all cases is slightly below
$T_{sp}$, a shoulder as a remnant of an incubation period in the
metastable regime is still visible, while a power--law--type decay
of $\Delta E(t)$ prevails for deeper quenches. A similar behavior
is found for $k_{||}(t)$, see Fig. 3 and also for $k_\perp(t)$
(not shown), which is always larger than $k_{||}(t)$. The
important observation is that within the timescales of our
simulations these power--laws change with $R$. As mentioned
already, at $R = 0$ where type--I walls have zero energy, an
anomalously slow growth is found, that can be represented by an
effective exponent $\alpha \simeq 1/4$ within an extended time
window up to our largest computing times. Upon introducing small
second--neighbor interactions with $R = 0.1$ we observe a
significantly faster growth. This becomes particularly evident by
comparing the data for the lowest temperature $T = 0.5\, T_0$ in
Figs. 2a and 3a with those in Figs. 2b and 3b. The slope of these
data corresponds to exponents still significantly smaller than
1/2. Further increase of $R$ to $R = 0.3$ (Figs. 2c and 3c) yields
exponents close to the ``normal'' behavior $\alpha = 1/2$,
pertaining here to $T \leq 0.7\, T_0$ and to a time regime
starting even below $10^2 $~MCS. No substantial change occurs when
going to $R = 0.5$, as shown in Figs. 2d and 3d. The steeper decay
near $t \sim 10^4 $~MCS can be traced back to finite size effects,
as $k_{||}^{-1}$ becomes of the order of the size $L = 128$ of the
simulation cell. These findings support the conclusion that the
extraordinary slow growth in the case of zero NNN--interactions
($R = 0$) arises from the existence of zero--energy, curvatureless
domain walls of type I, which are extremely stable. Within the
time window considered, such behavior appears representative for a
class of systems with $R$ small, where type--I walls have
sufficiently small, but non--zero energy.

Clearly, our results especially for $R \leq 0.1$ do not allow us
to draw any conclusion as to the exact asymptotics for quantities
displayed in Figs. 2 and 3. However, from the discussion at the
end of section 2 it appears that simulations up to $10^4 $~MCS
already exhaust the timescales relevant to experiments for
quenches sufficiently below the spinodal.

From the discussion of Fig.~1 it is clear that structure factors
$S_\alpha$ for a given $\alpha$ are affected by a 1--dimensional
(1--d) array of type--II walls parallel to the $\alpha$--direction
and a 2--d array of type--I walls perpendicular to it. Hence we
expect the structure factors $S_{||}$ and $S_\perp$ to obey
scaling laws for 1--d and 2--d systems, respectively \cite{bray}.
This is verified in Fig. 4 showing master curves for $S_{||}$ and
$S_\perp$ when scaled according to Porod's law in one and two
dimensions. This plot, applying to $R = 0.5$, is very similar to
analogous plots in \cite{paper2} for $R = 0$.

\begin{figure}[h]
\begin{center}
\epsfig{file=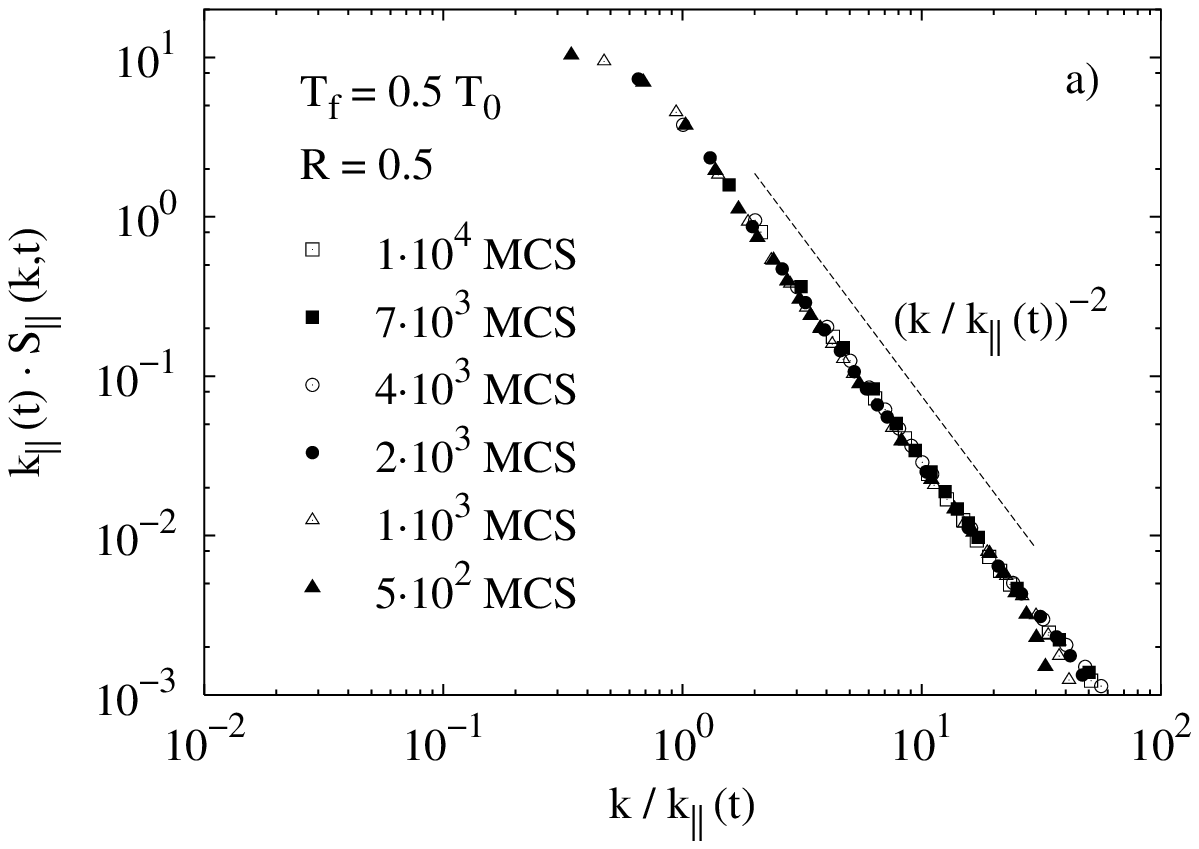, width=6.5cm} \epsfig{file=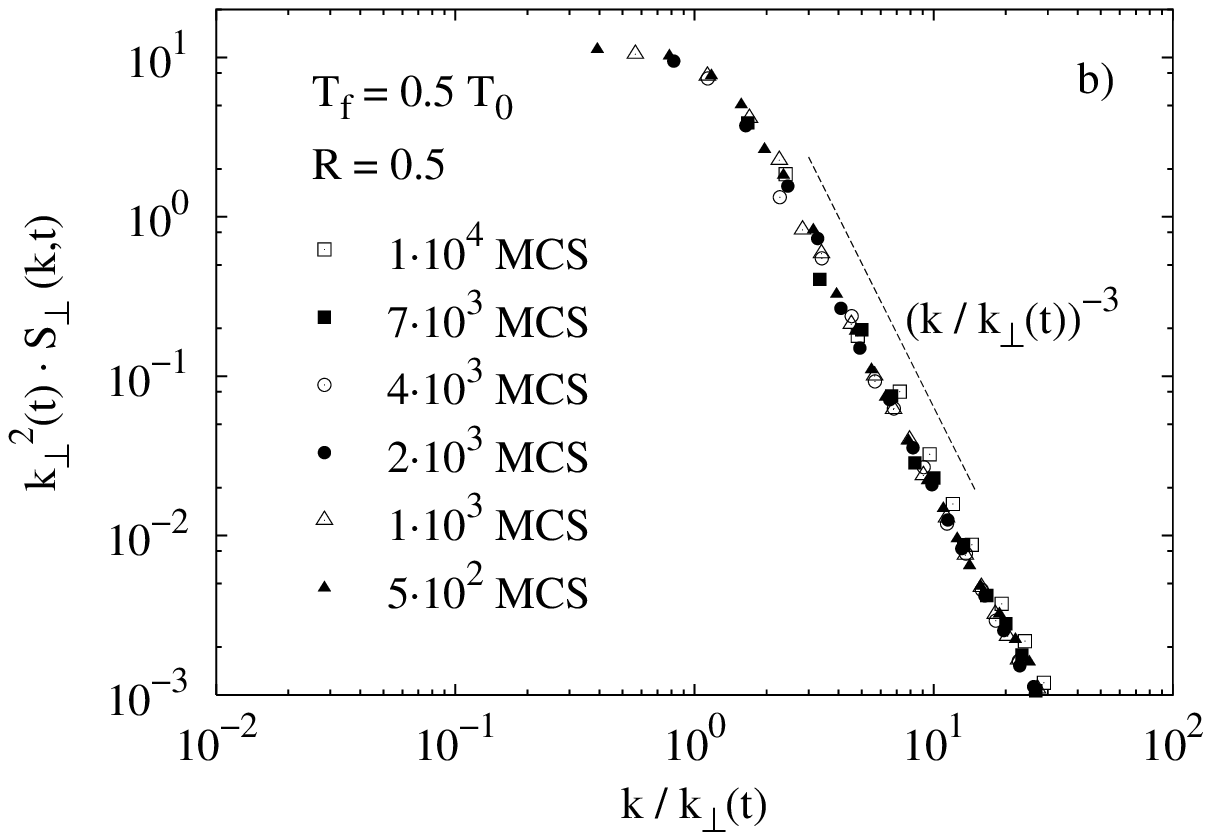,
width=6.5cm} \caption{Structure factors $S_{||}(k,t)$ a) and
$S_\perp (k,t)$ b) scaled according to Porod's law in $d=1$ and
$d=2$, respectively, at $T = 0.5\,T_0$ for $R = 0.5$.}
\end{center}
\end{figure}

\end{document}